\documentclass[british,english,aps,prb]{revtex4}
\usepackage[T1]{fontenc}
\usepackage[latin1]{inputenc}
\usepackage{graphicx}
\usepackage{amssymb}

\makeatletter

\providecommand{\boldsymbol}[1]{\mbox{\boldmath $#1$}}

\usepackage{babel}
\makeatother
\begin{document}

\title{Statistical Mechanics of Time Independent Non-Dissipative Nonequilibrium
States}

\date{\today}

\author{Stephen R. Williams}

\email{swilliams@rsc.anu.edu.au}

\affiliation{Research School of Chemistry, Australian National University, Canberra,
ACT 0200, Australia}

\author{Denis J. Evans}

\email{evans@rsc.anu.edu.au}

\affiliation{Research School of Chemistry, Australian National University, Canberra,
ACT 0200, Australia}

\begin{abstract}
We examine the question of whether the formal expressions of equilibrium
statistical mechanics can be applied to time independent non-dissipative
systems that are not in true thermodynamic equilibrium and are nonergodic.
By assuming the phase space may be divided into time independent,
locally ergodic domains, we argue that within such domains the relative
probabilities of microstates are given by the standard Boltzmann weights.
In contrast to previous energy landscape treatments, that have been
developed specifically for the glass transition, we do not impose
an a priori knowledge of the inter-domain population distribution.
Assuming that these domains are robust with respect to small changes
in thermodynamic state variables we derive a variety of fluctuation
formulae for these systems. We verify our theoretical results using
molecular dynamics simulations on a model glass forming system. Non-equilibrium
Transient Fluctuation Relations are derived for the fluctuations resulting
from a sudden finite change to the system's temperature or pressure
and these are shown to be consistent with the simulation results.
The necessary and sufficient conditions for these relations to be
valid are that the domains are internally populated by Boltzmann statistics
and that the domains are robust. The Transient Fluctuation Relations
thus provide an independent quantitative justification for the assumptions
used in our statistical mechanical treatment of these systems.
\end{abstract}
\maketitle

\section{Introduction}

The formal expressions of equilibrium statistical mechanics strictly
apply only to ergodic systems that are in thermodynamic equilibrium.
Thus these expressions only strictly apply to systems which are at
the global free energy minimum given the system Hamiltonian and the
macroscopic thermodynamic state variables (number of particles, temperature
and pressure or density). For such systems Gibbsian equilibrium statistical
mechanics provides an exact prescription for how to  calculate the
various thermodynamic quantities \cite{McQuarrie-Book}. However,
these prescriptions are routinely applied to systems that are not
in true thermodynamic equilibrium (for example to metastable liquids
\cite{Debenedetti-book}, glasses \cite{Debenedetti-Nat-01}, polymorphs
\cite{Sheng-NatMat-07} and allotropes). It is often observed empirically
that within experimental uncertainties many expressions for thermodynamic
quantities yield consistent results. In the present paper we provide
arguments for why many of the results of equilibrium statistical mechanics
can be applied to such time independent nondissipative nonequilibrium
systems. We also point out some of the limits inherent in the application
of the formulae of equilibrium statistical mechanics to such systems. 

We choose to study the isothermal isobaric ensemble \cite{Hansen-McDonald-book}
(externally regulated pressure and temperature). The methods and reasoning
we use here can be directly transferred to other ensembles such as
the canonical (fixed volume and externally regulated temperature).
The Gibbs free energy $G$, which is the thermodynamic potential for
the isothermal isobaric ensemble, is related to the partition function
$\Delta$ by the equation \begin{equation}
G(N,P_{0},T)=-k_{B}T\ln\Delta(N,P_{0},T),\label{G0}\end{equation}
and the partition function is given by the integral\begin{equation}
\Delta=\int\int_{D}dV\; d\boldsymbol{\Gamma}\exp[-\beta(H_{0}(\boldsymbol{\Gamma})+P_{0}V)],\label{Del0}\end{equation}
where $\boldsymbol{\Gamma}=(\boldsymbol{q},\boldsymbol{p})$ is the
phase space vector describing the coordinates $\boldsymbol{q}$ and
momenta $\boldsymbol{p}$, of all the $N$ particles in the system,
$P_{0}$ is the thermodynamic pressure, and $\beta=1/k_{B}T$ where
$k_{B}$ is Boltzmann's constant and $T$ is the temperature. The
integration domain $D$ provides limits for both integrals and extends
over all the available phase space $(\boldsymbol{\Gamma},V)$. This
is $\pm\infty$ for every component of the generalized momentum, $0\rightarrow\infty$
for the volume $V$, and over the volume for the Cartesian coordinates
of the particles. Since the system Hamiltonian $H_{0}(\boldsymbol{\Gamma},V)$
is single valued, so too is the partition function and in turn the
free energy. 

If we require the distribution function of a single thermodynamic
phase it is necessary that other phases do not contribute significantly
to the partition function. The full integration domain $D$ may include
states that are characteristic of crystalline states or fluids states.
In the thermodynamic limit this does not cause problems because, as
we shall see, the partition function  will be \emph{completely} dominated
by those microscopic domains that have the lowest free energy. However
the application of these formulae to allotropes or metastable systems
does present a problem. The standard equilibrium statistical mechanical
expressions for variables such as the enthalpy $I$, the average volume
$\left\langle V\right\rangle $ and second order quantities such as
the specific heat at constant pressure $c_{P}$ may all be computed
from a knowledge of the partition function Eq. \ref{Del0} or equivalently
the thermodynamic potential Eq. \ref{G0}. If other phases of lower
free energy exist this computation (from Eq. \ref{Del0} as written)
will strictly speaking be incorrect. 

It is well known that the formulae for thermal properties such as
entropy, free energy, temperature and specific heat do \emph{not}
hold for \emph{dissipative} nonequilibrium systems outside the linear
response regime \cite{Dorfman-book,evans-and-morriss-book}. In this
paper we examine the question of whether they are correct for any
\emph{nondissipative} nonequilibrium systems such as allotropes, metastable
systems or history dependent glasses. We provide a statistical mechanical
theory of time independent, nondissipative, nonequilibrium systems.
The theory is based on the fact that these systems are nonergodic
and individual sample systems comprise ergodic domains that do not
span all of phase space. We show that if these domains are robust
with respect to small changes in thermodynamic state variables, a
successful statistical mechanical treatment of these nonequilibrium
systems can be given. We provide direct evidence, from molecular dynamics
simulations on a model glass former, that the resulting statistical
mechanical formulae are satisfied within empirical errors. Finally
we provide an independent test of the two key elements of our theory:
Boltzmann weights within the phase space domains and the robustness
of those domains. It happens that these two elements are the necessary
and sufficient conditions for the application of the transient fluctuation
relation to finite thermodynamic quenches (in temperature or pressure)
for such systems \cite{Evans-Adv.-Phys.-02}. While the application
of thermodynamics to a single time averaged system is usually straightforward
the application to a ensemble, whose members may be locked in different
phase space domains, can require modification to the standard formulae.

In the case of glasses our treatment has some similarities with the
energy landscape approach of Stillinger and Weber \cite{Stillinger-PRA-82,Stillinger-sci-1984,Debenedetti-Nat-01}.
However there are significant differences; our treatment makes no
reference to the inherent structure and imposes no a priori knowledge
of the inter-domain relative population levels. The energy landscape
approach has been extended to account for the phenomena of ageing
or history dependence by the addition of a fictive parameter \cite{Sciortino-JSM-05}.
Sciortino has convincingly shown that the addition of a single fictive
parameter is inadequate to deal with glasses, which may have different
properties at the same temperature and pressure if they are prepared
by a different protocol (different history dependence) \cite{Sciortino-JSM-05}
and poses the challenge to recover a thermodynamic description {}``by
decomposing the ageing system into a collection of substates''. The
treatment we present here succeeds in doing just that by providing
a rigorous development of equilibrium statistical mechanics and thermodynamics
for ensembles of systems where the phase space breaks up into ensembles
of domains whose inter-domain dynamics is nonergodic and whose inter-domain
population levels may not be Boltzmann weighted.

\section{Conditions For Equilibrium}

A dynamical system in equilibrium has the properties that it is nondissipative
and that its macroscopic properties are time independent. Thus the
N-particle phase space distribution function, $f(\boldsymbol{\Gamma},V,t)$,
must be a time independent solution to the Liouville equation \cite{evans-and-morriss-book},\begin{equation}
\frac{\partial}{\partial t}f(\boldsymbol{\Gamma}^{\prime},t)=-\boldsymbol{\dot{\Gamma}}^{\prime}\cdot\nabla f(\boldsymbol{\Gamma}^{\prime},t)-f(\boldsymbol{\Gamma}^{\prime},t)\Lambda=0,\label{Liouville-Eq0}\end{equation}
where $\Lambda$ is the phase space compression factor \cite{evans-and-morriss-book}
obtained by taking the divergence of the equations of motion (see
Eq. \ref{Phase-space-compression}) and $\boldsymbol{\Gamma}^{\prime}$
is the extended phase space vector which consists of $\boldsymbol{\Gamma}$
and may include additional dynamical variables such as the volume
$V$. Since the system is assumed to be nondissipative both the ensemble
average $\left\langle \Lambda\right\rangle $ and the time average
$\overline{\Lambda}$ of the phase space compression factor (which
is directly proportional to the rate at which heat is exchanged with
the fictitious thermostat) are zero. The time independent solution
to Eq. \ref{Liouville-Eq0} depends on the details of the equations
of motion. Equilibrium solutions to Eq. \ref{Liouville-Eq0} for the
equations of motion, suitable for use in molecular dynamics simulations,
are compatible with Gibbsian equilibrium statistical mechanics \cite{evans-and-morriss-book}.

Microscopic expressions for mechanical properties like the pressure,
the internal energy, the enthalpy and the volume can be derived without
reference to Gibbsian statistical mechanics and indeed can be proved
to hold for nonequilibrium systems including nonequilibrium dissipative
systems. 

There are two ways in which the formulae derived from Gibbsian equilibrium
statistical mechanics can break down. The most obvious way is that
the relative weights of microstates may be non-Boltzmann and the exponential
factor $\exp[-\beta H_{0}(\boldsymbol{\Gamma})]$, may be replaced
by some other function (either the exponential function itself may
be modified as in Tsallis statistics \cite{Tsallis-physica-A-1998}
or the Hamiltonian may be modified to some new function $H_{0}(\boldsymbol{\Gamma})\rightarrow B(\boldsymbol{\Gamma},t)H_{0}(\boldsymbol{\Gamma})$).
In either circumstance the standard expressions for the thermal quantities
derived from equilibrium Gibbsian statistical mechanics will not be
valid. This certainly happens in dissipative nonequilibrium systems
where the distribution function is not a time independent solution
to Eq. \ref{Liouville-Eq0}.

In deterministic nonequilibrium steady states the phase space may
break down into ergodically separated domains (Each of which will
be fractal and of lower dimension than the ostensible phase space
dimension. This is a consequence of dissipation.) However for these
steady states, the domains are always exquisitely sensitive to macroscopic
thermodynamic parameters since they are strange fractal attractors
\cite{Hoover-book}. Often a deterministic \emph{nonequilibrium} steady
state approaches a unique fractal attractor. As time progresses the
distribution function collapses ever closer to (but never reaching)
the steady state attractor.

The second way that these expressions may fail is that the system
may become nonergodic. In this case three things happen. a) Most obviously
time averages no longer equal full (domain $D$) ensemble averages.
b) If we take an initial microstate the subsequent phase space trajectory
will span some phase space domain $D_{\alpha}$ where the initial
phase is labeled $\boldsymbol{\Gamma}_{\alpha}^{\prime}(0)$. In this
case for nondissipative nonequilibrium systems where the domains are
robust (i.e. small changes in thermodynamic state parameters, to leading
order do not change the domain) the standard equations of equilibrium
statistical mechanics may continue to be valid but in a slightly modified
form. We will examine this in some detail below. c) Given robust domains
the population densities between each domain may well depend on the
history of the system. The \emph{macroscopic} history can be expected
to condition the ensemble's set of initial \emph{microstates} $\{\boldsymbol{\Gamma}_{\alpha}^{\prime}(0)\}$
from which the macroscopic material is formed. This in turn can be
expected to condition the set of nonergodic domains $\{ D_{\alpha}\}$
that characterize the ensemble. For a macroscopic sample spanning
a single ergodic domain $D_{\alpha}$, the free energy $G_{\alpha}$
then satisfies only a local extrema principle and thus looses much
of its thermodynamic meaning.

\section{Theory and Methods}

\subsection{Equations of Motion}

We use the constant pressure Nos\'e-Hoover equations of motion by
combining the Nos\'e-Hoover feedback mechanism with the so-called
SLLOD or DOLLS equations of motion, which are equivalent for dilation.
It is known that these equations of motion do not produce artifacts
in the systems linear response to an external field and that to leading
order the effect on the dynamical correlation functions is at most
$\mathcal{O}(1/N)$, where $N$ is the number of particles \cite{evans-and-morriss-book}.
The equations of motion are,\begin{eqnarray}
\dot{\mathbf{q}}_{i} & = & \frac{\mathbf{p_{i}}}{m}+\alpha_{V}\mathbf{q_{i}}\nonumber \\
\dot{\mathbf{p}}_{i} & = & \mathbf{F}_{i}-\alpha_{V}\mathbf{p}_{i}-\alpha_{T}\mathbf{p_{i}}\nonumber \\
\dot{\alpha}_{V} & = & \left(\frac{V(t)}{Nk_{B}T}\left(P(t)-P_{0}\right)+\frac{1}{N}\right)/\tau_{V}^{2}\nonumber \\
\dot{\alpha}_{T} & = & \left(\frac{\sum_{i=1}^{N}\boldsymbol{p}_{i}\cdot\boldsymbol{p}_{i}}{3mNk_{B}T}-1+\frac{1}{N}\right)/\tau_{T}^{2}\nonumber \\
\dot{V} & = & 3\alpha_{V}V,\label{EOM}\end{eqnarray}
where $\mathbf{q}_{i}$ is the position, $\mathbf{p}_{i}$ is the
momentum and $\mathbf{F}_{i}$ is the force on the $i^{th}$ particle,
$m$ is the particle mass, $\tau_{V}$ is the barostat time constant,
$\tau_{T}$ is the thermostat time constant, $T$ is the input temperature,
$P_{0}$ is the input (thermodynamic) pressure and the instantaneous
(mechanical) pressure is given by $P(t)=(\sum_{i=1}^{N}\boldsymbol{p}_{i}\cdot\boldsymbol{p}_{i}/m+\sum_{i=1}^{N}\boldsymbol{F}_{i}\cdot\boldsymbol{q}_{i})/3V$.
Because these equations of motion have additional dynamical variables
the extended phase space vector is $\boldsymbol{\Gamma}^{\prime}=(\boldsymbol{\Gamma},V,\alpha_{V},\alpha_{T})$.
In order to obtain the equilibrium distribution function we first
define the Hamiltonian, in the absence of any external fields, dilation
$\alpha_{V}(t)=0$ or thermostats $\alpha_{T}(t)=0$, as $H_{0}=\Phi+\frac{1}{2}\sum_{i=1}^{N}\boldsymbol{p}_{i}\cdot\boldsymbol{p}_{i}/m$,
where $\Phi$ is the total inter-particle potential energy. To proceed
further we identify the extended Hamiltonian as $H_{E}=H_{0}+\frac{3}{2}N\alpha_{T}^{2}\tau_{T}^{2}k_{B}T+\frac{3}{2}N\alpha_{V}^{2}\tau_{V}^{2}k_{B}T$
and then obtain the phase space compression factor\begin{eqnarray}
\Lambda & \equiv & \nabla\cdot\dot{\boldsymbol{\Gamma}^{\prime}}=\sum_{i=1}^{N}\sum_{\gamma=1}^{3}\frac{\partial\dot{q}_{i,\gamma}}{\partial q_{i,\gamma}}+\sum_{i=1}^{N}\sum_{\gamma=1}^{3}\frac{\partial\dot{p}_{i,\gamma}}{\partial p_{i,\gamma}}+\frac{\partial\dot{V}}{\partial V}+\frac{\partial\dot{\alpha}_{V}}{\partial\alpha_{V}}+\frac{\partial\dot{\alpha}_{T}}{\partial\alpha_{T}}\nonumber \\
 & = & \beta(\dot{H}_{E}+P_{0}\dot{V}),\label{Phase-space-compression}\end{eqnarray}
where the index $\gamma$ sums over the components of the Cartesian
position and momentum vectors. Using the Heisenberg streaming representation
(rather than the more usual Schr\"odinger representation Eq. \ref{Liouville-Eq0})
of the Liouville equation\begin{equation}
\frac{d}{dt}\ln\left[f(\boldsymbol{\Gamma}^{\prime}(t),t)\right]=-\Lambda\left(\boldsymbol{\Gamma}^{\prime}(t)\right),\label{Liouville-streaming}\end{equation}
we can obtain the particular time independent solution for the distribution
function\begin{equation}
f(\boldsymbol{\Gamma}^{\prime})\propto\exp(-\beta I_{0})\exp(-\frac{3}{2}N(\alpha_{T}^{2}\tau_{T}^{2}+\alpha_{V}^{2}\tau_{V}^{2})),\label{dist0}\end{equation}
where $I_{0}(t)=H_{0}(t)+P_{0}V(t)$ is the instantaneous enthalpy.
The second exponential on the RHS of Eq. \ref{dist0} with $\alpha_{V}$
and $\alpha_{T}$ in the argument, which has no dependence on the
input temperature $T$ or the input pressure $P_{0}$, is statistically
independent from the rest of the distribution function, which is the
standard equilibrium isothermal isobaric distribution. We can normalize
Eq. \ref{dist0} by integrating over all space to obtain the thermodynamic
equilibrium distribution function, \begin{equation}
f(\boldsymbol{\Gamma}^{\prime})=\frac{3}{2}N\frac{\tau_{V}\tau_{T}}{\pi}\exp(-\frac{3}{2}N(\alpha_{T}^{2}\tau_{T}^{2}+\alpha_{V}^{2}\tau_{V}^{2}))f_{0}(\boldsymbol{\Gamma},V),\label{dist1}\end{equation}
where the standard isothermal isobaric distribution function is\begin{equation}
f_{0}(\boldsymbol{\Gamma},V)=\frac{exp(-\beta(H_{0}+P_{0}V))}{\int_{0}^{\infty}dV\int_{D}d\boldsymbol{\Gamma}exp(-\beta(H_{0}+P_{0}V))}.\label{dist-standard}\end{equation}
It should be emphasized that the derivation of Eq. \ref{dist0} says
nothing about the existence or otherwise of any domains. These must
be considered when normalizing Eq. \ref{dist0} and thus Eqs. \ref{dist1}
\& \ref{dist-standard} are only valid in thermodynamic equilibrium.
If we wish to use Eq. \ref{dist0} outside thermodynamic equilibrium
we must consider domains.

We can also use the so called SLLOD equations of motion \cite{evans-and-morriss-book}
to apply strain rate controlled Couette flow (planar shear) to our
equations of motion. The necessary modifications to the first two
lines of Eq. \ref{EOM} result in\begin{eqnarray}
\dot{\mathbf{q}}_{i} & = & \frac{\mathbf{p}_{i}}{m}+\alpha_{V}\mathbf{q}_{i}+\mathbf{i}\dot{\gamma}q_{yi}\nonumber \\
\dot{\mathbf{p}}_{i} & = & \mathbf{F}_{i}-\alpha_{V}\mathbf{p}_{i}-\alpha_{T}\mathbf{p}_{i}-\mathbf{i}\dot{\gamma}p_{yi},\label{SLLOD}\end{eqnarray}
where $\dot{\gamma}$ is the strain rate and the last three lines
of Eq. \ref{EOM} remain unchanged.

\subsection{Equilibrium Statistical Mechanics in a Single Domain}

As we have stated in the introduction, the full phase space domain
includes phase points from many different thermodynamic phases (gases,
liquids and crystals). In the thermodynamic limit this does not cause
problems. To understand this suppose we can label microstates to be
in either of two possible thermodynamic phases 1 or 2 bound by two
phase space domains $D_{1}$ and $D_{2}$. By assumption we are not
presently considering the possibility of co-existence. The system
is assumed to be ergodic: atoms in one thermodynamic phase can in
time, transform into the other phase. Assume the two thermodynamic
phases have different free energies: $G_{1}$ is the Gibbs free energy
of the first phase and $G_{2}$ is that of the second phase. For a
sufficiently large $N$ the free energy Eq. \ref{G0} is an extensive
variable. We may thus express the partition function as the sum of
contributions from the two phases\begin{eqnarray}
\Delta & = & e^{-\beta G_{1}}+e^{-\beta G_{2}}\nonumber \\
 & = & e^{-\beta Ng_{1}}+e^{-\beta Ng_{2}},\label{G-sums}\end{eqnarray}
where the lower case $g$ on the second line is used to represent
the intensive free energies which do not change with system size $N$
in the thermodynamic limit. If $g_{1}$ is less than $g_{2}$ then
in the thermodynamic limit, $N\rightarrow\infty$, the only significant
contribution to the partition function $\Delta$ will be due to the
{}``equilibrium'' phase namely phase 1. Thus although the free energy
defined in Eq. \ref{Del0}, is given by an integral over all of phase
space $D$, in the thermodynamic limit this integral can be approximated
to arbitrary precision, as an integral over the domain that includes
the most stable phase. Suppose $D_{1}$ includes only crystalline
phases and $D_{2}$ includes only amorphous phases and further suppose
a particular crystalline phase has a lower free energy that any amorphous
phase. According to Eq. \ref{Del0}, we should calculate the free
energy by integrating over all crystalline and all amorphous phases.
In practice in the thermodynamic limit we can compute the free energy
to arbitrary accuracy by integrating Eq. \ref{Del0}, only over that
part of phase space within which the thermodynamically stable state
resides.

If we consider a nonergodic system that according to different preparative
protocols can be formed in either phase 1 or phase 2. After preparation,
because the system is nonergodic both phases are kinetically stable
indefinitely. By restricting the phase space integrals for the free
energy to those domains that contain the kinetically stable phase
we can compute the free energy of that phase. However, although it
may be possible to formally assign free energies to nonergodic systems,
these free energies clearly fail to satisfy any global extremum principle.
As we will show these partition functions can be used formally to
yield first and second order thermodynamic quantities by numerical
differentiation. The metastable domain is a subset of the thermodynamic
equilibrium domain which contains all possible atom positions including
ones belonging to the metastable phase.

Within a single domain the system is, by construction, ergodic. Thus
for almost all microstates $\boldsymbol{\Gamma}_{\alpha}^{\prime}(0)$$\in$$D_{\alpha}$
, ensemble averages, of some variable $B$, $\left\langle B\right\rangle $,
equal time averages $\overline{B}$, for phase space trajectories
that start at time zero, \begin{equation}
\left\langle B\right\rangle _{\alpha}=\overline{B}_{\alpha}\equiv\lim_{t\rightarrow\infty}\frac{1}{t}\int_{0}^{t}ds\; B(\boldsymbol{\Gamma}^{\prime}(s);\boldsymbol{\Gamma}_{\alpha}^{\prime}(0)).\label{time-av-def}\end{equation}
Microscopic expressions for mechanical variables may be used as a
test of ergodicity in nondissipative systems which are out of equilibrium.
(Note a nondissipative system does not on average exchange heat with
any thermal reservoir with which it has been in contact for a long
time.) In the case of metastable fluids or allotropes we may introduce
a single restricted domain and by construction the system remains
ergodic within this domain. 

A \emph{gedanken} experiment can be used to justify the Boltzmann
weighting and the applicability of the Zeroth Law of Thermodynamics
for such systems. Consider a double well potential with an inner and
outer potential well. If the barrier between the inner and outer wells
is much greater than $k_{B}T$, so that over the duration of observation
(which is much greater than any relaxation time in the ergodically
restricted subsystem) no particles cross the barrier, then the system
considered as a double well system, will be, by construction, nonergodic.
For systems composed of particles that are solely found in the inner
potential well, our hypotheses are that the distribution of states
in the inner well will be given by a Boltzmann distribution taken
over the inner domain only and that if such a system is in thermal
contact with a body in true thermodynamic equilibrium, then the temperature
of the ergodically restricted system must equal that of the system
in true thermodynamic equilibrium. We can justify these hypotheses
by considering a fictitious system that only has the inner potential
well and in which the potential function is positive infinity for
all separations that are greater than the inner well (this includes
the position of the outer well). In accord with Gibbsian statistical
mechanics the distribution of states is canonical over this (single
well) potential. Furthermore the Zeroth Law of thermodynamics will
apply to this single well system. Now if we dynamically generate the
outer well, all the particles locked inside the inner well cannot
{}``know'' that the outer well has been formed so their dynamics
will be completely unchanged by the time dependent generation of the
new outer well. The generation of an inaccessible outer well will
not alter the distribution of states in the inner well nor will it
cause any flow of heat to the equilibrium heat bath surrounding the
system. This provides a compelling physical justification for our
domain hypotheses over a single ergodic sub domain of phase space.

In order to recover many of the basic relationships of Gibbsian statistical
mechanics it is also necessary that the system appears to be in dynamical
equilibrium, i.e. $\mid f(\boldsymbol{\Gamma}_{\alpha},t)-f(\boldsymbol{\Gamma}_{\alpha},t+\tau_{o})\mid<\varepsilon,\forall\;\boldsymbol{\Gamma}_{\alpha},{\in D}_{\alpha}$,
for some small $\varepsilon$, over the longest observation time $\tau_{o}$.
We use the definition of the partition function Eq. \ref{Del0} as
before but now the domain $D_{\alpha}$ in the integral is over a
single contiguous hypervolume in the configuration space of the generalized
position coordinate $\boldsymbol{q}$ \textbf{\large }and volume $V$.
The domain over the generalized momentum $\boldsymbol{p}$ and multipliers
$\alpha_{V}$ and $\alpha_{T}$ remains unchanged. We then obtain
the Gibbs free energy by use of Eqs. \ref{G0} \& \ref{Del0}. Thus
far all we have altered is our definition of the domain. In changing
the definition of the domain we have opened a potential problem for
Gibbsian statistical mechanics. If we change the temperature or the
pressure of the system the domain may also change. If the domain changes
this may make a contribution to the derivatives of the partition function,
Eq. \ref{Del0}, and the direct connection with the standard outcomes
of macroscopic thermodynamics will be lost. Thus the domains need
to be robust with respect to changes in thermodynamic state variables.

There are three means by which a system could have robust domains.
The first and most obvious is that the domain does not change when
the pressure or the temperature changes, $\partial D_{\alpha}(X,Y)/\partial X=0$,
where $X$ is a thermodynamic state variable and $Y$ is the other
thermodynamic state variables. When we lower the temperature only
the inverse temperature $\beta$ in Eq. \ref{dist0} changes and when
we change the pressure only the parameter $P_{0}$ changes. If the
domain's boundary is determined by a surface on which $I_{0}(\boldsymbol{\Gamma}^{\prime})$
always has a very high value it will remain unchanged under infinitesimal
changes in $P_{0}$ or $\beta$. We will refer to a surface domain
that doesn't change with the state variables as completely robust.
The second way is that the distribution function is always identically
zero on the domain boundary,

\begin{equation}
f_{0}(\boldsymbol{\Gamma},V)=0\;\;\forall\;\;\boldsymbol{\Gamma}\in S_{\alpha},\label{boundary0}\end{equation}
where $S_{\alpha}$ is the surface of the domain $D_{\alpha}$ . Because
the domain is contiguous (required for it to be ergodic) it must have
a single connected surface. Such a domain will be robust. The third
way the domain can be robust is less restrictive and allows for the
possibility that the domain does change when the thermodynamic variables
are changed substantially. If $\delta X$ is an infinitesimal change
in a thermodynamic state variable then,

\begin{equation}
\delta D_{\alpha}(X+\delta X,Y)=\delta D_{\alpha}(X,Y)+\mathcal{O}(\delta X)^{n}\label{eq:domainvariation}\end{equation}
where $Y$ denotes the other thermodynamic state variables, we require
that $n\geq2$ for first order thermodynamic property formulae to
be correct $n\geq3$ for second order property formulae to be correct
etc. Obviously this third way will be satisfied in the first two cases
as well.

Later in the paper we will introduce an independent test of domain
robustness. However, if a system was not robust then we would expect
that small changes in the state variables would change the macroscopic
properties of the sample permanently - it would be as though the preparation
history of the sample was continuing even for small changes in the
state variable. Quite obviously if we produce huge changes in the
state variables we will of course permanently change the properties
of the system because we permanently deform the ergodic domain. Experience
shows however that very many nondissipative nonequilibrium systems
are quite robust with respect to small changes in state variables.
All that is required for fluctuation formulae for first second and
third order thermodynamic quantities to be valid, is that the domains
be unchanged, to first second or third order, by infinitesimal changes
in the state variables. Obviously a robust domain is an ideal construct.
However on the typical time scale of interest, which is usually orders
of magnitude less than the time scale on which the system will change
to a new phase of lower free energy, this can be a very good approximation. 

We are now able to recover most of the standard results of Gibbsian
equilibrium statistical mechanics. For example we may calculate the
enthalpy $\left\langle I\right\rangle _{\alpha}$ from the partition
function $\Delta$, Eq. \ref{Del0}, as\begin{eqnarray}
\left\langle I\right\rangle _{\alpha} & = & k_{B}T^{2}\left(\frac{\partial ln\Delta_{D_{\alpha}}}{\partial T}\right)\nonumber \\
 & = & \frac{\int\int_{D_{\alpha}}dV\, d\boldsymbol{\Gamma}I_{0}(\boldsymbol{\Gamma},V)exp(-\beta I_{0}(\boldsymbol{\Gamma},V))}{\int\int_{D_{\alpha}}dV\, d\boldsymbol{\Gamma}exp(-\beta I_{0}(\boldsymbol{\Gamma},V))}=\overline{I}_{\alpha}.\label{Enthalpy}\end{eqnarray}
Here we are considering an ensemble of systems which occupy a single
ergodic domain $D_{\alpha}$. Since this domain is self ergodic the
ensemble average is equal to the corresponding time average. 

The term on the RHS of the second line is obviously the average value
of the instantaneous enthalpy, $I_{0}(\boldsymbol{\Gamma},V)=H_{0}(\boldsymbol{\Gamma},V)+P_{0}V$,
obtained by using the equilibrium distribution function, Eq. \ref{dist-standard}
or equivalently Eq. \ref{dist1} with the integration limits restricted
to the domain $D_{\alpha}$, where $P_{0}$ is the externally set
thermodynamic pressure. We can also obtain expressions for the average
volume $\left\langle V\right\rangle $ and the constant pressure specific
heat $c_{P}$ by taking the appropriate derivatives of  the partition
function Eq. \ref{Del0}. In other ensembles we can use the same procedure
to find other variables e.g. the internal energy, the average pressure
and the constant volume specific heat in the canonical (\emph{N},\emph{V},\emph{T})
ensemble.

An important outcome is that this description remains compatible with
macroscopic thermodynamics. Here the Gibbs free energy is defined
as,\begin{equation}
G\equiv U-TS+P_{0}\left\langle V\right\rangle ,\label{Gibbs thermo}\end{equation}
where $U=\left\langle H_{0}\right\rangle $ is the internal energy
and $S$ is the entropy. If we take the derivative of Eq. \ref{Gibbs thermo}
with respect to one of the isobaric isothermal ensembles conjugate
variables $(N,P_{0},T)$ while keeping the others fixed we obtain\begin{eqnarray}
\left(\frac{dG}{dT}\right)_{N,P_{0}} & = & -S\nonumber \\
\left(\frac{dG}{dP_{0}}\right)_{N,T} & = & \left\langle V\right\rangle .\label{Gibbs conj deriv}\end{eqnarray}
We now write down the microscopic equilibrium equation for the Gibbs
entropy\begin{equation}
S=-k_{B}\int\int_{D_{\alpha}}dV\, d\boldsymbol{\Gamma}f_{0}(\boldsymbol{\Gamma},V)\ln f_{0}(\boldsymbol{\Gamma},V).\label{Gibbs Entropy}\end{equation}
It is an easy matter to show that Eqs. \ref{G0}, \ref{Del0} \& \ref{Gibbs Entropy}
are consistent with the two derivatives given in Eq. \ref{Gibbs conj deriv}.
Given our condition of a robust boundary we thus have a form of Gibbsian
statistical mechanics for metastable states which remains in agreement
with macroscopic thermodynamics.

\subsection{Multiple Domains and Nonergodicity}

We now wish to consider an ensemble of systems which is prepared from
an initial ergodic (usually high temperature) equilibrium ensemble.
There is some protocol $\mathcal{P}$, which involves a temperature
quench or a sharp pressure increase etc, which breaks the ensemble
into a set of sub-ensembles characterized by different macroscopic
properties. After the protocol $\mathcal{P}$, has been executed we
allow all the ensemble members to relax to states which are macroscopically
time independent - to within experimental tolerances. We assume that
the ensemble can be classified into a set of sub-ensembles $\{\alpha,\alpha=1,N_{D}\}$
whose macroscopic properties take on $N_{D}$ distinct sets of values.
For the longest observation times available a macroscopic system classified
as an $\alpha$ system is not observed to transform into a $\beta$
system, and vice versa. The full ensemble of systems is thus non-ergodic.
However, in each individual sub-ensemble, say sub-ensemble $\alpha$,
the constituent members are ergodic (by construction). Thus we can
partition the full phase space into a set of domains, $\{ D_{\alpha}\}$. 

From the arguments given above (in section B), after the relaxation
of initial transients,  we expect to observe a Boltzmann distribution
of states within an individual domain  which is therefore independent
of the quench protocol. However the distribution between domains cannot
be expected to be Boltzmann distributed and will instead be dependent
on the quench protocol. Within a given ensemble the proportion of
ensemble members ultimately found in domain $D_{\alpha}$ is given
by a weight $w_{\alpha}(\mathcal{P})$ which is subject to the constraint
\begin{equation}
\sum_{\alpha=1}^{N_{D}}w_{\alpha}=1.\label{w-constraint}\end{equation}
We can calculate the full \emph{ensemble} average of some macroscopic
property $B$ as,\begin{eqnarray}
\left\langle B\right\rangle  & = & \sum_{\alpha=1}^{N_{D}}w_{\alpha}\frac{\int\int_{D_{\alpha}}dV\, d\boldsymbol{\Gamma}B(\mathbf{\Gamma},V)\,\exp(-\beta I_{0})}{\int\int_{D_{\alpha}}dV\, d\boldsymbol{\Gamma}\exp(-\beta I_{0})}.\label{multi-D-average}\end{eqnarray}
Since the full ensemble of states is nonergodic the phase space breaks
up into disjoint domains which in themselves are ergodic. Thus each
domain may be identified by any point in phase space, $(\boldsymbol{\Gamma},V)$,
that is a member of it so the subscript $\alpha$ is a function of
the phase vector $\alpha(\boldsymbol{\Gamma},V)$ allowing the following
expression for the distribution function\begin{equation}
f(\boldsymbol{\Gamma},V)=\sum_{\alpha=1}^{N_{D}}w_{\alpha}s(\boldsymbol{\Gamma},D_{\alpha})f_{\alpha}(\mathbf{\Gamma},V)\label{multi-D-dist}\end{equation}
where $s(\boldsymbol{\Gamma},D_{\alpha})=1$ if $\mathbf{\Gamma}\epsilon\, D_{\alpha}$
and $s(\boldsymbol{\Gamma},D_{\alpha})=0$ otherwise, and \begin{equation}
f_{\alpha}(\boldsymbol{\Gamma},V)=\frac{\exp(-\beta I_{0}(\boldsymbol{\Gamma},V))}{\int_{0}^{\infty}dV\int_{D_{\alpha}}d\boldsymbol{\Gamma}\exp(-\beta I_{0}(\boldsymbol{\Gamma},V))}.\label{multi-D-dist-raw}\end{equation}
The entropy is given by $S=-k_{B}\int_{0}^{\infty}dV\int_{D}d\boldsymbol{\Gamma}f\ln(f)$
and using Eq. \ref{multi-D-dist} we have the following expressions
for the multidomain entropy,\begin{equation}
S=-k_{B}\sum_{\alpha=1}^{N_{D}}w_{\alpha}\left[\int_{0}^{\infty}dV\int_{D_{\alpha}}d\boldsymbol{\Gamma}f_{\alpha}\ln(f_{\alpha})+\ln(w_{\alpha})\right]=\sum_{\alpha=1}^{N_{D}}w_{\alpha}S_{\alpha}-k_{B}\sum_{\alpha=1}^{N_{D}}w_{\alpha}\ln(w_{\alpha}).\label{S-multi-D}\end{equation}
The term $-\sum_{\alpha=1}^{N_{D}}k_{B}w_{\alpha}\ln(w_{\alpha})\equiv S_{D}$
is the inter-domain entropy, which is maximized by an even distribution
of ensemble members over all domains, while $S_{\alpha}$ is the intra-entropy
of domain $\alpha$ considered as a single \emph{N}-particle system. 

If we substitute Eq. \ref{multi-D-dist-raw} into Eqs. \ref{S-multi-D}
we find that,

\begin{eqnarray}
S & = & S_{D}+T^{-1}\sum_{\alpha=1}^{N_{D}}w_{\alpha}\left\langle I_{0}\right\rangle _{\alpha}+k_{B}\sum_{\alpha=1}^{N_{D}}w_{\alpha}ln\int_{0}^{\infty}dV\int_{D_{\alpha}}d\boldsymbol{\Gamma}\exp(-\beta I_{0}(\boldsymbol{\Gamma},V))\nonumber \\
 & = & S_{D}+T^{-1}\left\langle I\right\rangle +k_{B}\sum_{\alpha=1}^{N_{D}}w_{\alpha}ln\int_{0}^{\infty}dV\int_{D_{\alpha}}d\boldsymbol{\Gamma}\exp(-\beta I_{0}(\boldsymbol{\Gamma},V)),\label{S-multi-D-G}\end{eqnarray}
where $\left\langle B\right\rangle _{\alpha}=\int\int_{D_{\alpha}}dV\, d\mathbf{\Gamma}\, B(\mathbf{\Gamma})f_{\alpha}(\mathbf{\Gamma},V)$.
Combining Eq. \ref{Gibbs thermo} with Eq. \ref{S-multi-D-G} we obtain
the following expression for the Gibbs free energy

\begin{equation}
G=-k_{B}T\sum_{\alpha=1}^{N_{D}}w_{\alpha}\left[\ln\int\int_{D_{\alpha}}dV\, d\boldsymbol{\Gamma}\exp(-\beta I_{0})-\ln(w_{\alpha})\right]=\sum_{\alpha=1}^{N_{D}}w_{\alpha}G_{\alpha}-S_{D}T.\label{G-multi-D}\end{equation}
It is easy to verify that if we hold the local domain weights fixed
and then vary the temperature or the pressure that Eqs. \ref{S-multi-D}
\& \ref{G-multi-D} are compatible with Eqs. \ref{Gibbs conj deriv}.
This means that if we have a fixed number of robust domains, whose
population levels or weights are non-Boltzmann distributed, Eqs. \ref{G-multi-D}
\& \ref{S-multi-D} provide a direct microscopic link to standard
macroscopic thermodynamics. On the extremely long time scale the weighting
functions $w_{\alpha}$ may vary and the system will tend towards
the direction where the free energy Eq. \ref{G-multi-D} is reduced.
Without the inter-domain entropy term, $S_{D}$, Eq. \ref{G-multi-D}
would be minimized when the domain with the lowest free energy has
all the ensemble members in it. It turns out that Eq. \ref{G-multi-D}
is minimized when all the domain weights are Boltzmann distributed,
ie when\begin{equation}
w_{\alpha}=\frac{\int\int_{D_{\alpha}}dV\, d\boldsymbol{\Gamma}exp(-\beta I_{0})}{\sum_{\beta=1}^{N_{D}}\int\int_{D_{\beta}}dV\, d\boldsymbol{\Gamma}exp(-\beta I_{0})}.\label{weights Boltzmann}\end{equation}
Here (i.e. upon obeying Eq. \ref{weights Boltzmann}) the entropy
and free energy given by Eqs. \ref{S-multi-D} \& \ref{G-multi-D}
coincide with the standard equilibrium expressions so the free energy
must be a minimum. To prove this we use Eq. \ref{G-multi-D} and we
remove the first weight $w_{1}=1-\sum_{\alpha=2}^{N_{D}}{\displaystyle {\displaystyle w_{\alpha}}}$,
so that the constraint Eq. \ref{w-constraint} is respected while
the remaining weights are independent. This means that the free energy
can be written as $G(1-{\displaystyle \sum_{\alpha=2}^{N_{D}}{\displaystyle w_{\alpha}}},w_{2},w_{3},...w_{N_{D}})$.
The constrained partial derivatives are then \begin{eqnarray}
\left.\frac{\partial G}{\partial w_{\alpha}}\right|_{c} & = & \frac{\partial G}{\partial w_{1}}\frac{\partial w_{1}}{\partial w_{\alpha}}+\frac{\partial G}{\partial w_{\alpha}}=-\frac{\partial G}{\partial w_{1}}+\frac{\partial G}{\partial w_{\alpha}},\,\alpha\geq2\nonumber \\
 & = & -\frac{\partial G}{\partial w_{1}}+\frac{\partial G}{\partial w_{\alpha}},\,\alpha\geq2,\nonumber \\
 & = & -G_{1}-kTln(w_{1})-k_{B}T+G_{\alpha}+k_{B}Tln(w_{\alpha})+k_{B}T,\,\alpha\geq2.\label{dGdw}\end{eqnarray}
Using the fact that for Boltzmann weights, Eq. \ref{weights Boltzmann},\begin{equation}
w_{\alpha}=exp[\beta(G_{eq}-G_{\alpha})]\label{weights-Boltz-2}\end{equation}
where $G_{eq}=-k_{B}T\ln\sum_{\alpha=0}^{N_{D}}\int\int_{D_{\alpha}}dV\, d\mathbf{\Gamma}\exp(-\beta I_{0})$
is the equilibrium free energy, we find that at equilibrium,

\begin{equation}
\frac{\partial G}{\partial w_{\alpha}}\mid_{c}=0,\,\alpha\geq2.\label{dGdw-min}\end{equation}
It remains to prove that this is indeed a minimum. Using the same
approach for treating the constraint, we continue making the first
weight a function of all the others, and obtain\begin{equation}
\left.\frac{\partial^{2}G}{\partial w_{\alpha}\partial w_{\gamma}}\right|_{c}=\frac{\partial^{2}G}{\partial w_{1}^{2}}-\frac{\partial^{2}G}{\partial w_{\gamma}\partial w_{\alpha}}-\frac{\partial^{2}G}{\partial w_{\gamma}\partial w_{1}}-\frac{\partial^{2}G}{\partial w_{1}\partial w_{\alpha}}.\label{Hessian-1}\end{equation}
Using the Boltzmann weights it is easy to show that\begin{eqnarray}
\frac{\partial^{2}G}{\partial w_{1}^{2}} & = & \frac{kT}{w_{1}}=k_{B}Texp[\beta(G_{1}-G_{eq})]\nonumber \\
\frac{\partial^{2}G}{\partial w_{1}\partial w_{\alpha}} & = & \frac{\partial^{2}G}{\partial w_{\gamma}\partial w_{1}}=0\nonumber \\
\frac{\partial^{2}G}{\partial w_{\gamma}\partial w_{\alpha}} & = & \delta_{\gamma\alpha}\frac{k_{B}T}{w_{\alpha}}=\delta_{\gamma\alpha}k_{B}Texp[\beta(G_{\alpha}-G_{eq})].\label{Hessian-2}\end{eqnarray}
From these results it is easy to see that the Hessian matrix $\left.\frac{\partial^{2}G}{\partial w_{\alpha}\partial w_{\gamma}}\right|_{c}$
is positive definite, thus we have proved the free energy to be a
minimum for the case of equilibrium.

We can use a knowledge of the multiple domain thermodynamic potential,
Eq. \ref{G-multi-D}, to compute averages. As an example we consider
the average enthalpy again,\begin{equation}
\left\langle I_{0}\right\rangle =-k_{B}T^{2}\,\left.\frac{\partial\,\beta G}{\partial T}\right|_{w_{\alpha},P_{0},N}.\label{Enthalpy-multi-d}\end{equation}
Eq. \ref{Enthalpy-multi-d} can easily be derived from Eq. \ref{Gibbs thermo}
and is in essence the same as the first line of Eq. \ref{Enthalpy}.
It is straightforward to see that upon using Eq. \ref{G-multi-D}
to calculate the derivative in Eq. \ref{Enthalpy-multi-d} one obtains
the average enthalpy as given by Eq. \ref{multi-D-average}. One can
do the same for other quantities such as the specific heat etc.

\subsection{Application to Molecular Dynamics Simulation}

We are now in a position to test various outcomes, using computer
simulation, which may be derived by drawing on the previous material.
If we start with an ensemble of systems which are initially in equilibrium
at temperature $T=T_{0}$ and then at time $t=0$ we quench them by
setting $T=T_{1}$ we can solve the Liouville equation Eq. \ref{Liouville-streaming}
to obtain\begin{eqnarray}
f(\boldsymbol{\Gamma}(t),\alpha_{V}(t),\alpha_{T}(t),t) & = & f\left(\boldsymbol{\Gamma}(0),\alpha_{V}(0),\alpha_{T}(0),0\right)\nonumber \\
 & \times & \exp\left[\beta_{1}(I_{0}(0)-I_{0}(t))\right]\label{quench dist}\end{eqnarray}
where $\beta_{1}=1/k_{B}T_{1}$. This nonequilibrium distribution
function, valid for $t>0$, explicitly requires the solution of the
equations of motion and is of limited utility. However it allows the
identification of a formal condition to identify the amount of time,
which must elapse after the quench, before Eqs. \ref{dist0}, \ref{dist-standard},
\ref{multi-D-average} or \ref{multi-D-dist} can be applied to the
ensemble. That is the quantity $I_{0}(t)=H_{0}(t)+P_{0}V(t)$ must
be statistically independent of $I_{0}(0)$. Thus we are interested
in the correlation function \begin{eqnarray}
C_{1}\left(I_{0}(t),I_{0}(0)\right) & = & \frac{\left\langle I_{0}(t)I_{0}(0)\right\rangle -\left\langle I_{0}(t)\right\rangle \left\langle I_{0}(0)\right\rangle }{C_{1,0}}\label{correlation1}\end{eqnarray}
where \begin{equation}
C_{1,0}=\sqrt{\left(\left\langle I_{0}(t)^{2}\right\rangle -\left\langle I_{0}(t)\right\rangle ^{2}\right)\left(\left\langle I_{0}(0)^{2}\right\rangle -\left\langle I_{0}(0)\right\rangle ^{2}\right)}.\label{cor1-norm}\end{equation}
This function will equal 1 for a perfectly correlated system, -1 for
a perfectly anticorrelated system and 0 for an uncorrelated system.
When we consider an ensemble of systems, occupying the various domains
to different degrees, we see that Eq. \ref{correlation1} may not
decay to zero given the trajectories are unable to leave their domains.
If the transients, due to the quench, fully decay Eq. \ref{correlation1}
will decay to zero and Eqs. \ref{dist0} \& \ref{dist-standard} will
become valid for the ensemble. 

If the correlation function, Eq. \ref{correlation1} decays to a plateau
then we may have a situation where Eqs. \ref{multi-D-average} and
\ref{multi-D-dist} are valid. It may be that the material can still
slowly age, due to processes, that occur on a time scale which is
longer than the one we are monitoring. For a glass we expect that
correlation function Eq. \ref{correlation1} will not fully decay
on a reasonable time scale. If we give the system time to age, such
that it appears to be time translationaly invariant, and then compute
the following correlation function,\begin{equation}
C_{2}\left(I_{0}(\tau),I_{0}(0)\right)=\frac{\sum_{\alpha=0}^{N_{t}}\left\langle I_{0}(t)I_{0}(0)\right\rangle _{\alpha}-\left\langle I_{0}(t)\right\rangle _{\alpha}\left\langle I_{0}(0)\right\rangle _{\alpha}}{C_{2,0}}\label{correlation2}\end{equation}
where\begin{equation}
C_{2,0}=\sum_{\alpha=0}^{N_{t}}\sqrt{\left(\left\langle I_{0}(t)^{2}\right\rangle _{\alpha}-\left\langle I_{0}(t)\right\rangle _{\alpha}^{2}\right)\left(\left\langle I_{0}(0)^{2}\right\rangle _{\alpha}-\left\langle I_{0}(0)\right\rangle _{\alpha}^{2}\right)},\label{cor2-norm}\end{equation}
we may observe a full decay. If this occurs Eqs. \ref{multi-D-average}
and \ref{multi-D-dist} will be valid. To compute this correlation
function $N_{t}$ trajectories are produced and for each of these
the averages, $\left\langle \bar{B}\right\rangle _{\alpha}$, where
$B$ is an arbitrary dynamical variable, appearing in Eq. \ref{correlation2}
are approximately obtained by time averaging. When the system is ergodic
and time translationaly invariant these two correlation functions,
Eqs. \ref{correlation1} \& \ref{correlation2}, will give the same
result. However for a nonergodic system $C_{2}$ will decay to zero
on a reasonable time scale while $C_{1}$ will not. Rather $C_{1}$
may decay to some plateau on a reasonable time scale and then decay
on a much slower time scale. The preceding sections then rest on this
clear separation of time scales in the correlation function Eq. \ref{correlation1}.
For metastable fluids and allotropes this separation is so extreme
that we probably cannot observe, even the early stages of, the later
slow decay on any reasonable experimental time scale. Further for
these systems there will be only a single domain and thus they appear
ergodic. For glasses some signs of the later decay can often be observed,
however it is still very much slower than the initial decay. In the
field of glassy dynamics the initial decay is often called the $\beta$
decay and the slower long time decay is often called the $\alpha$
decay. As the glass is further aged this second stage decay is observed
to slow down dramatically while the early decay does not change very
much \cite{vanMegen-PRE-98}.

To allow the hypothesis of local phase space domains to be tested
we will now introduce several relations whose derivation draws upon
the equilibrium distribution function Eq. \ref{dist-standard}. We
will also discuss the effect of the phase space domains on these relations.

First we introduce the configurational temperature\begin{equation}
k_{B}T=-\frac{\left\langle \boldsymbol{F}\cdot\boldsymbol{F}\right\rangle }{\left\langle \nabla\cdot\boldsymbol{F}\right\rangle }\label{config temp}\end{equation}
where $\boldsymbol{F}$ is a $3N$ dimensional vector representing
the inter-particle forces on each atom $\boldsymbol{F}=-\mathcal{\nabla}\Phi$.
This relation is easily derived from Eq. \ref{dist-standard} (set
$B$ to $\nabla\cdot\boldsymbol{F}$, integrate by parts and drop
the boundary terms) and will remain valid for nonergodic systems where
Eq. \ref{weights Boltzmann} is not obeyed. The relation involves
the spatial derivative of the force, which is not zero at the cutoff
radius for the potential we use in our simulations (see below). This
along with finite size effects can result in a small disagreement
between Eq. \ref{config temp} and the kinetic temperature for our
system when in equilibrium.

Equilibrium fluctuation formulae may be easily derived from Eq. \ref{multi-D-average},
see ref. \cite{Hansen-McDonald-book,McQuarrie-Book} for some examples
of how this is done. We consider how the average enthalpy changes
with the temperature at constant pressure,\begin{equation}
C_{P_{0}}=\left\langle V\right\rangle c_{P_{0}}=\left.\frac{d}{dT}\right|_{N,P_{0}}\left\langle I_{0}\right\rangle =\frac{\beta}{T}\left[\left\langle I_{0}^{2}\right\rangle -\left\langle I_{0}\right\rangle ^{2}\right]\label{Cp-regular}\end{equation}
where $c_{P_{0}}$ is the constant pressure specific heat. The calculation
of the RHS of Eq. \ref{Cp-regular} using the ensemble average given
by Eq. \ref{multi-D-average} is subtle. If we simply calculate $\left\langle I_{0}\right\rangle $
and $\left\langle I_{0}^{2}\right\rangle $ with the use of Eq. \ref{multi-D-average}
and then plug the results into Eq. \ref{Cp-regular} we obtain, what
we will refer to as, a \emph{single domain average} which does not
give us the correct change in the average enthalpy for a history dependent
equilibrium ensemble. This is because the average $\left\langle I_{0}^{2}\right\rangle $
superimposes across the different domains while the quantity $\left\langle I_{0}\right\rangle ^{2}$
contains spurious cross terms. If we derive the heat capacity by taking
the second derivative of the Gibbs free energy for the multiple domain
distribution function, Eqs. \ref{G-multi-D} \& \ref{Enthalpy-multi-d},
with respect to the temperature, \begin{equation}
C_{P_{0}}=-\left.\frac{\partial}{\partial T}\right|_{N,P_{0},w_{\alpha}}\, k_{B}T^{2}\left.\frac{\partial\,\beta G}{\partial T}\right|_{N,P_{0},w_{\alpha}},\label{Cp-multi-thermo}\end{equation}
we obtain the following 

\begin{equation}
C_{P_{0}}=\left\langle V\right\rangle c_{P_{0}}=\frac{\beta}{T}\sum_{\alpha=1}^{N_{D}}w_{\alpha}\left[\left\langle I_{0}^{2}\right\rangle _{\alpha}-\left\langle I_{0}\right\rangle _{\alpha}^{2}\right],\label{Cp-concat}\end{equation}
where the quantity $\left\langle I_{0}^{2}\right\rangle _{\alpha}-\left\langle I_{0}\right\rangle _{\alpha}^{2}$
is obtained for each domain separately (here $\left\langle \ldots\right\rangle _{\alpha}$
represents an average taken where all ensemble members are in the
$\alpha^{th}$ domain). We will refer to this as a \emph{multidomain
average} which is consistent with the nonergodic statistical mechanics
and thermodynamics that we have introduced here. It is obvious that
both a single and multiple domain average will give the same result
in the case of thermodynamic equilibrium and metastable equilibrium
(single domain). The transition from the single domain average producing
the correct result to an anomalous result is symptomatic of an ergodic
to a history dependent nonergodic transition. If we consider a large
macroscopic system (the super system) to be made of $N_{s}$ independent
subsystems the multidomain average remains self-consistent. To see
this we apply Eq. \ref{Cp-concat} to fluctuations in the super system
and then we inquire how this relates to fluctuations in the subsystem.
The enthalpy of one instance of the super system will be given by
$I_{s}=\sum_{\alpha=1}^{N_{s}}I_{\alpha}$. In principle an ensemble
of super systems can be prepared by applying the same history dependent
macroscopic protocol to all members of this ensemble. Due to the statistical
independence of the subsystems, upon taking an ensemble average of
super systems, we have $\left\langle I_{\alpha}I_{\beta}\right\rangle _{S}=\left\langle I_{\alpha}\right\rangle _{S}\left\langle I_{\beta}\right\rangle _{S}$
for all $\alpha\neq\beta$. Here the average $\left\langle \ldots\right\rangle _{S}$
is taken over the ensemble of super systems and the $\alpha^{th}$
subsystem in each super system is identified by its location. It is
then easy to show that the specific heat obtained from the ensemble
average Eq. \ref{Cp-concat} of the super system is equivalent to
that obtained from the subsystem due to the two quantities $\left\langle I_{s}^{2}\right\rangle _{S}=\left\langle \left(\sum_{\alpha=1}^{N_{s}}I_{\alpha}\right)^{2}\right\rangle _{S}$
and $\left\langle I_{s}\right\rangle _{S}^{2}=\left\langle \sum_{\alpha=1}^{N_{s}}I_{\alpha}\right\rangle _{S}^{2}$
possessing identical cross terms which cancel each other out (as a
result of the independence of the subsystems) upon applying Eq \ref{Cp-concat}. 

If we ignore finite size effects, due to assuming the equivalence
of ensembles, the constant volume specific heat is related to the
constant pressure specific heat by the equation\begin{equation}
C_{V}=\left\langle V\right\rangle c_{V}=C_{P}-P\left.\frac{dV}{dT}\right|_{T}-\left(\left.\frac{\partial H}{\partial P}\right|_{T}\left/\,\frac{\partial V}{\partial P}\right|_{T}\right)\;\left.\frac{\partial V}{\partial T}\right|_{P}.\label{Cv}\end{equation}
We may also obtain an expression for the constant volume specific
heat $c_{V}$ by deriving equilibrium fluctuation formula for each
of the derivatives appearing in Eq. \ref{Cv} in lieu of directly
measuring them. We may then obtain a single domain expression for
$c_{V}$ which does not work for the history dependent glass and also
a correctly weighted ensemble average (multidomain average) which
does. This is completely analogous to what has been shown in detail
for $c_{P}$. As the equations are unwieldy, and their derivation
(given an understanding of the $c_{P}$ case) is straightforward,
we will not reproduce them here.

\subsection{Test of Domain Robustness: Transient Fluctuation Theorem}

The application of the Evans-Searles Transient Fluctuation Theorem
\cite{Evans-PRL-93-ECM2,Evans-Adv.-Phys.-02,Searles-AJC-04} to the
systems treated in this paper provides a sharp test of the assumptions
used to develop the theory given in this paper. The Theorem describes
a time reversal symmetry satisfied by a generalized entropy production,
namely the so-called dissipation function. The precise mathematical
definition of this function requires a knowledge of the dynamics and
also of the initial distribution function. The three necessary and
sufficient conditions for the Fluctuation theorem to be valid are
that the initial distribution is known (here we assume the distribution
is Boltzmann weighted over some initial domain of phase space), that
the dynamics is time reversible (all the equations of motion used
here are time reversible) and lastly that the system satisfies the
condition known as ergodic consistency. When applied to the systems
studied here this requires that the phase space domains should be
robust with respect to the sudden changes imposed on the system and
that the number of inter-domain transitions remain negligible on the
time over which the theory is applied. If any one of these three conditions
fails then the Theorem cannot be applied to the system and the corresponding
fluctuation relation will not be satisfied\cite{Evans-Adv.-Phys.-02}.

We can use the fluctuation theorem to obtain relations for how the
system responds upon suddenly changing the input temperature or pressure
for a system, which is initially in equilibrium as specified by Eq.
\ref{dist0} or \ref{dist1}. Firstly we consider a change in the
pressure, while holding the temperature fixed, by changing the input
variable $P_{0}$ in Eq. \ref{EOM} (thermodynamic pressure) to $P_{0}=P_{2}$
at time $t=0$ for a system initially in equilibrium with $P_{0}=P_{1}$.
The probability density $p(\Delta V=A)$ of observing a change in
volume of $\Delta V(t)=V(t)-V(0)$ relative to a change of equal magnitude
but opposite sign is then given by\begin{equation}
\frac{p\left(\Delta V(t)=A\right)}{p\left(\Delta V(t)=-A\right)}=\exp\left(\beta(P_{2}-P_{1})A\right).\label{TFT-P}\end{equation}
To derive this expression we have had to assume that the intra-domain
populations are Boltzmann distributed according to Eq. \ref{dist0}.
Ergodic consistency requires that for any initial phase space point
$\mathbf{\Gamma}(0)$ that can be initially observed with nonzero
probability, there is a nonzero probability of \emph{initially} observing
the time reversal map $M^{T}$of the end point $\Gamma(t)$, (ie $\forall\:\mathbf{\Gamma}(0)$
such that $f(\mathbf{\Gamma}(0),0)\neq0,$$\:$$f(M^{T}(\mathbf{\Gamma}(t)),0)\neq0$).
This condition obviously requires that the phase space domains remain
robust and the number of inter-domain transitions remain negligible
for at least a time $t$, after the pressure (or temperature) quench.

If we sample all or our initial $t=0$, $P_{0}=P_{1}$ configurations
from the one trajectory which remains locked in a single domain even
after the quench, we expect Eq. \ref{TFT-P} to be valid. If we prepare
an ensemble of initial configurations using the same protocol we still
expect Eq. \ref{TFT-P} to remain valid even with different domain
weightings $w_{i}$, as defined in Eqs. \ref{multi-D-average} \&
\ref{multi-D-dist}, provided the domains are robust over the time
$t$ appearing in Eq. \ref{TFT-P}. Note that suddenly reducing the
pressure by a very large amount could result in a breakdown of the
robustness condition. Eq. \ref{TFT-P} may be partially summed to
obtain what is referred to as the integrated fluctuation theorem \begin{equation}
\frac{p\left(\Delta V(t)>0\right)}{p\left(\Delta V(t)<0\right)}=\left\langle \exp\left(\beta(P_{2}-P_{1})\Delta V\right)\right\rangle _{\Delta V<0}.\label{IFT-P}\end{equation}

For the case where we change the input temperature $T$ in Eq. \ref{EOM}
while holding the input pressure $P_{0}$ fixed we obtain a relation
for fluctuations in the extended instantaneous enthalpy $I_{E}(t)=H_{E}(t)+P_{0}V(t)$.
We start with a system initially in equilibrium at temperature $T=1/(k_{B}\beta_{1})$
and we then subject it to a temperature quench by changing the input
temperature in Eq. \ref{EOM} to $T=1/(k_{B}\beta_{2})$, at time
$t=0$, while holding the input pressure fixed. The probability density
$p\left(\Delta I_{E}(t)=A\right)$ of observing a change in instantaneous
enthalpy $\Delta I_{E}(t)=I_{E}(t)-I_{E}(0)$ relative to a change
of equal magnitude but opposite sign is then given by\begin{equation}
\frac{p\left(\Delta I_{E}(t)=A\right)}{p\left(\Delta I_{E}(t)=-A\right)}=\exp\left((\beta_{1}-\beta_{2})A\right).\label{TFT-T}\end{equation}
Note that if we suddenly increase the temperature by a very large
amount we could expect to violate the robustness or the negligible
inter-domain transition condition. In common with Eq. \ref{TFT-P}
we expect that this expression will be valid when all initial configurations
are sampled from a single common domain and also when sampled from
multiple arbitrarily populated domains under the assumption that the
domains are robust and the number of transitions are negligible over
time $t$. This equation may also be partially summed to obtain\begin{equation}
\frac{p\left(\Delta I_{E}(t)>0\right)}{p\left(\Delta I_{E}(t)<0\right)}=\left\langle \exp\left((\beta_{2}-\beta_{1})\Delta I_{E}\right)\right\rangle _{\Delta I_{E}<0}.\label{IFT-T}\end{equation}

\section{Simulation Details}

For our simulations we use a variation on the Kob Andersen glass former
\cite{Kob-PRE-95a} featuring a purely repulsive potential \cite{Williams-PRL-06}.
The pairwise additive potential is\begin{eqnarray}
u_{ij}(r_{ij})=4\epsilon_{\alpha\beta} & \left[\left(\frac{\sigma_{\alpha\beta}}{r_{ij}}\right)^{12}-\left(\frac{\sigma_{\alpha\beta}}{r_{ij}}\right)^{6}+\frac{1}{4}\right] & \;\;\;\forall\;\;\; r_{ij}<\sqrt[6]{2}\,\sigma_{\alpha\beta}\nonumber \\
u_{ij}(r_{ij})=0\;\;\;\;\;\, &  & \;\;\;\forall\;\;\; r_{ij}>\sqrt[6]{2}\,\sigma_{\alpha\beta},\label{potential}\end{eqnarray}
where the species identities of particles $i$ and $j$, either $A$
or $B$, are denoted by the subscripts $\alpha$ and $\beta$. The
energy parameters are set $\epsilon_{BB}=0.5\,\epsilon_{AA}$, $\epsilon_{AB}=1.5\,\epsilon_{AA}$
and the particle interaction distances $\sigma_{BB}=0.88\,\sigma_{AA}$,
$\sigma_{AB}=0.8\,\sigma_{AA}$. The energy unit is $\epsilon_{AA}$,
the length unit is $\sigma_{AA}$ and the time unit is $\sqrt{m\,\sigma_{AA}^{2}/\epsilon_{AA}}$
with both species having the same mass $m$. The composition is set
at $X=N_{B}/N_{A}=0.2$, the number of particles are $N=N_{A}+N_{B}=108$,
the pressure is set to $P_{0}=14\;\epsilon_{AA}/\sigma^{3}$ and the
temperature unit is $\epsilon_{AA}/k_{B}$. The time constants are
set at $\tau_{V}=5\sqrt{N}$ and $\tau_{T}=\sqrt{N}$. Note that the
energy parameters are slightly different to the potential we used
in ref. \cite{Williams-PRL-06}. The equations of motion were integrated
using a fourth order Runge-Kutta method \cite{Butcher-ANM-96}. The
time step used was $dt=0.002$ and sometimes $dt=0.004$ for very
low temperatures. 

From previous work on binary mixtures we know the basic reason why
this system is vary reluctant to crystallize \cite{Williams-PRE-01a,Fernandez-PRE-03,Fernandez-JCP-04,Fernandez-JPCB-2004a}.
The chosen nonadditivity of the species \emph{A}-\emph{B} interaction
makes the mixture extremely miscible; consider the present value of
$\sigma_{AB}=0.8\,\sigma_{AA}$ relative to the additive value of
$\sigma_{AB}=0.94\,\sigma_{AA}$. This effect dominates over the choice
of the energy parameters. Due to this extreme miscibility the relatively
large composition fluctuations necessary, about the average composition
of $X=0.2$, to form the crystal phases (either the pure species \emph{A,}
$X=0$, FCC crystal or the binary, $X=0.5$, CsCl crystal) are strongly
suppressed and crystallization is strongly frustrated.

\begin{figure}
\includegraphics[scale=0.9]{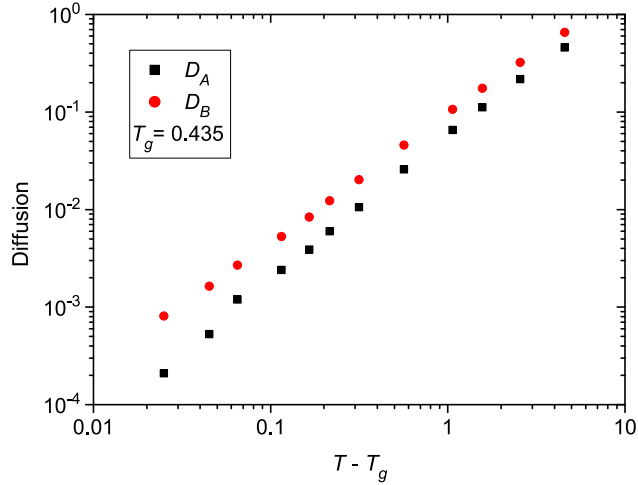}

\caption{A logarithmic plot of the self diffusion coefficient for both species
\emph{A} and species \emph{B} particles as a function of the separation
parameter $T-T_{g}$ with $T_{g}=0.435$.\label{fig-1}}
\end{figure}

We identify the nominal glass transition by calculating the diffusion
coefficient as a function of temperature. This is shown for both species
in Fig. \ref{fig-1} on a logarithmic plot demonstrating how the diffusion
coefficient approaches zero critically, $D_{\alpha}(T)\propto(T-T_{g})^{b}$
where $b$ is the critical exponent, with a nominal glass transition
temperature of $T_{g}=0.435$. It would be, perhaps, more customary
to obtain a nominal glass transition temperature by analyzing the
critical divergence of the viscosity. Given that the Stokes Einstein
relation is strongly violated upon approaching the glass transition
one might be concerned that this would give a very different result.
However the violation of the Stokes Einstein relation can largely
be attributed to the exponent $b$ being different between the viscosity
and the diffusion coefficient rather than the nominal glass transition
temperature $T_{g}$ \cite{Puertas-JPCM-05}.

\section{Results and Discussion}

\begin{figure}
\includegraphics[scale=0.9]{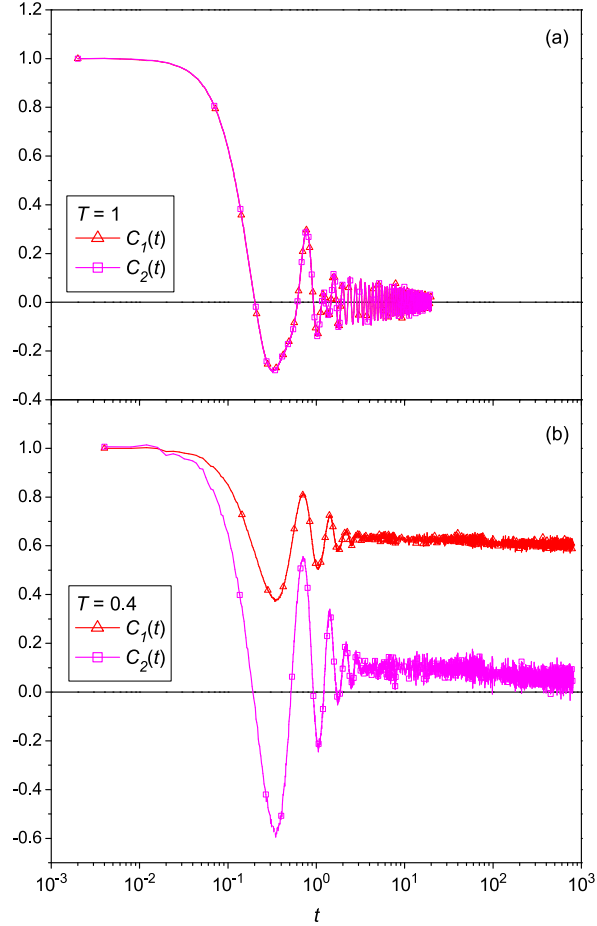}

\caption{a) The instantaneous enthalpy correlation functions as defined by
Eqs. \ref{correlation1} \& \ref{correlation2} as a function of logarithmic
time for the temperature $T=1$. The calculation of the function was
started after the fluid had been given time to equilibrate. The strong
agreement between the two correlation functions is indicative of ergodicity.\protect \\
b) The instantaneous enthalpy correlation functions as defined by
Eqs. \ref{correlation1} \& \ref{correlation2} as a function of logarithmic
time for the temperature $T=0.4$. The calculation of the correlation
function was started at various times after the quench, all approximately
at $t=8\times10^{5}$, in an attempt to age the system. The difference
between the correlation functions is indicative of nonergodicity.
Notice that $C_{2}$ reaches a near full decay between $t=1\;\&\;10$,
while $C_{1}$ reaches a non-decaying plateau. \label{fig-2}}
\end{figure}

The correlation functions given in Eqs. \ref{correlation1} \& \ref{correlation2}
were calculated from ensembles of $100$ independent simulations at
the two temperatures given in Fig. \ref{fig-2} ($T=1$ and $T=0.4$).
In all cases the systems were subject to an instantaneous quench,
from an initial equilibrium at $T=5$, by changing the value of the
input temperature in Eq. \ref{EOM}. The system was then run for a
significant time, in the case of the glass ensemble $\tau_{age}>8\times10^{5}$,
in an attempt to age it. Of course the longest time that can be accessed
in a molecular dynamics simulation is rather short, and so the system
is not very well aged, but we are still able to meaningfully treat
it as a time invariant state. Each of the $100$ independent simulations
was interpreted as being stuck in its own domain $D_{\alpha}$ and
the correlation functions were calculated for each of these domains
using time averaging. The time averaging was approximately $100$
times longer than the longest time $t=800$ that the correlation functions
were calculated out to. Obviously in the limit of an infinite number
of independent simulations and the case where the domains are robust
 we will obtain the exact multidomain average given by Eq. \ref{multi-D-average}.
We assume our limited ensemble of simulations is representative of
this. The data from each domain (independent simulation) was then
used to obtain the correlation functions Eqs.\ref{correlation1} \&
\ref{correlation2} as seen in Fig. \ref{fig-2}. %
 At the higher temperature $T=1$ it can be seen that the two functions
are equivalent demonstrating how the system is ergodic. It can also
be seen that the correlation function has decayed on a time scale
of $t\sim10$ which is therefore (by Eq. \ref{quench dist}) the time
scale on which the ensemble becomes accurately represented by Eq.
\ref{dist-standard} with only one domain $D$ which does not necessarily
extend over all phase space. The oscillations, which can be seen in
the correlation function, are due to both ringing in the feedback
mechanisms of Eq. \ref{EOM} and the frequency dependent storage component
of the bulk viscosity. The statistical uncertainty in the correlation
function becomes larger than these oscillations somewhere between
a time of $t=1\;\&\;10$. If we constructed an experiment where the
pressure was regulated by a piston and a spring, the correlation functions,
Eqs.\ref{correlation1} \& \ref{correlation2}, would depend on the
details of the piston and spring parameters in a similar way to the
simulations dependence on the details of the feed back mechanism. 

When the system is quenched to the lower temperatures ($T=0.4$) ergodicity
is lost and we obtain a glass. The complete decay of the first correlation
function, $C_{1}$, Eq. \ref{correlation1} may no longer occur because
the individual trajectories remain stuck in local domains which have
different average values for $\left\langle I_{0}\right\rangle _{\alpha}$.
This is similar to the much studied density correlation function \cite{Kob-PRE-95b}
(the intermediate scattering function) which decays to a finite plateau
for a glass or more generally a solid material. On the other hand
there is nothing to stop the second correlation function, $C_{2}$,
Eq. \ref{correlation2}, from fully decaying when the system is nonergodic.
If the second correlation function fully decays whilst the first is
only able to decay to a plateau then we have a situation where Eqs.
\ref{multi-D-dist} \& \ref{multi-D-average} are valid as can be
seen from Eq. \ref{quench dist}. In Fig. \ref{fig-2} it can be seen
that $C_{1}$ does indeed fail to decay while $C_{2}$ comes very
close to fully decaying at the time where $C_{1}$ reaches the plateau.
The reason $C_{2}$ doesn't fully decay here is due to the fact that
the  inter-domain transition rates, while small, are not exactly zero.
If the system had been aged more extensively this problem would be
significantly reduced. This effect is exacerbated by the time averaging,
used to form the averages for each trajectory, being two orders of
magnitude longer than the longest time the correlation function was
calculated to. The effect of the state slowly evolving due to  finite
inter-domain transition rates is too small to seriously compromise
the modeling of the system as obeying Eqs. \ref{multi-D-average}
\& \ref{multi-D-dist} and thus we have obtained direct evidence for
the validity of these equations. The height of the plateau for $C_{1}$
will depend on the history of the system, i.e. the protocol used to
prepare the ensemble.

\begin{figure}
\includegraphics[scale=0.9]{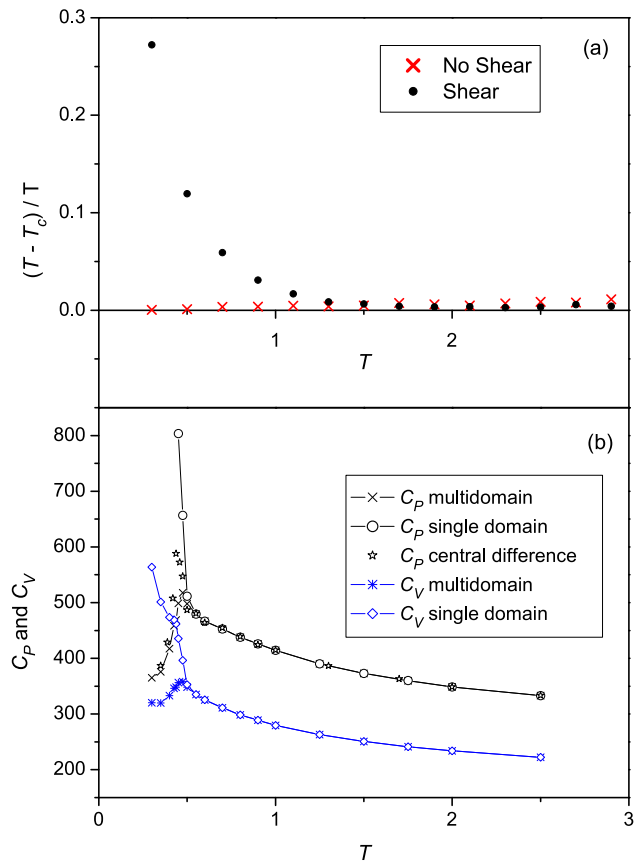}

\caption{a) The relative difference between the kinetic temperature $T$ (controlled
directly by the Nos\'e Hoover thermostat) and the configurational
temperature given by Eq. \ref{config temp} as a function of the kinetic
temperature. The results for the system undergoing Couette flow (Shear)
are for a constant strain rate of $\dot{\gamma}=0.5$. \protect \\
b) The heat capacity calculated using the equilibrium fluctuation
formula by the single domain averaging method Eq. \ref{Cp-regular}
and the multidomain method Eq. \ref{Cp-concat}. Also shown is data
obtained by numerically differentiating the enthalpy by central difference.
At the temperatures above the peak the three types of averages give
very similar results. Also shown is equivalent equilibrium fluctuation
formula data for the constant volume specific heat Eq. \ref{Cv}.
\label{fig-3}}
\end{figure}

We move on to a comparison between the kinetic temperature and the
configurational temperature, the results of which may be seen in Fig.
\ref{fig-3}a). %
 The input temperature ranges from $T=3$ to $T=0.3$. Also shown
are results for the system, undergoing constant planar shear, Eq.
\ref{SLLOD}, with a strain rate of $\dot{\gamma}=0.5$. At the higher
temperatures we see a very small relative discrepancy between the
two types of temperatures, which we attribute to the discontinuity
in the first spatial derivative of the inter-particle force at the
cutoff radius and to finite size effects. These effects appear to
diminish a little at lower temperatures. For temperatures above T
= 1.5 the chosen strain rate has no significant effect on the configurational
temperature indicating that our system is in the linear response domain
\cite{Williams-PRL-06}. At the lowest temperatures, well below the
glass transition temperature, we observe good agreement between the
configurational and input temperatures for the system without shear.
This provides further evidence of our assertion that the system obeys
Boltzmann statistics in the glass Eqs. \ref{multi-D-average} \& \ref{multi-D-dist}.
However, at low temperatures, the system that is undergoing shear
shows an increasing relative discrepancy between the two temperatures.
At low temperatures the system leaves the linear response domain \cite{Williams-PRL-06}
demonstrating the fundamental difference between the nonequilibrium
distribution of the history dependent glassy state and that of a strongly
driven steady state. If we drive the system hard enough, at any given
temperature, we can always make a disagreement between the two types
of temperature due to the steady state no longer being accurately
represented by a Boltzmann distribution i.e. due to a break down in
local thermodynamic equilibrium. When the system is not driven by
an external field we have been unable to observe any difference in
the two temperatures by deeply supercooling a glass forming mixture
apart from the initial transient decay immediately following the quench,
which falls off surprisingly rapidly. 

In Fig \ref{fig-3}b) results are presented for the heat capacities
(the specific heat multiplied by the volume) at both constant pressure
$C_{P}$ and constant temperature $C_{V}$. Details of the protocol
used to obtain this data is given in the end note \cite{endnote-1}.
The estimates from the multidomain average are compared with those
from the single domain average. The results from the multidomain averages
exhibit the well-known peak, which is a signature of the onset of
the glass transition, and has been observed directly by calorimetry
in many experiments on real glass forming materials \cite{Debenedetti-book}.
The temperature, where the peak is observed, depends on the history
of the system. No peak is observed for the single domain averages
which continue to increase as the temperature is lowered. While the
two methods for forming averages give the same results at temperatures
above the peak, they diverge at temperatures below the peak. It is
the multidomain average that gives results consistent with the actual
calorimetric behavior of the system. This may be seen in the figure
by comparing the data which has been computed by numerically differentiating
the enthalpy using central difference. At the peak neither the central
difference (due to rapid rate of change) or the multidomain average
data (due to a lack of domain robustness) are reliable and they show
significant differences. However below the peak they once again show
quantitative agreement providing strong evidence that the domains
are robust in this region. %
\begin{figure}
\includegraphics[scale=0.9]{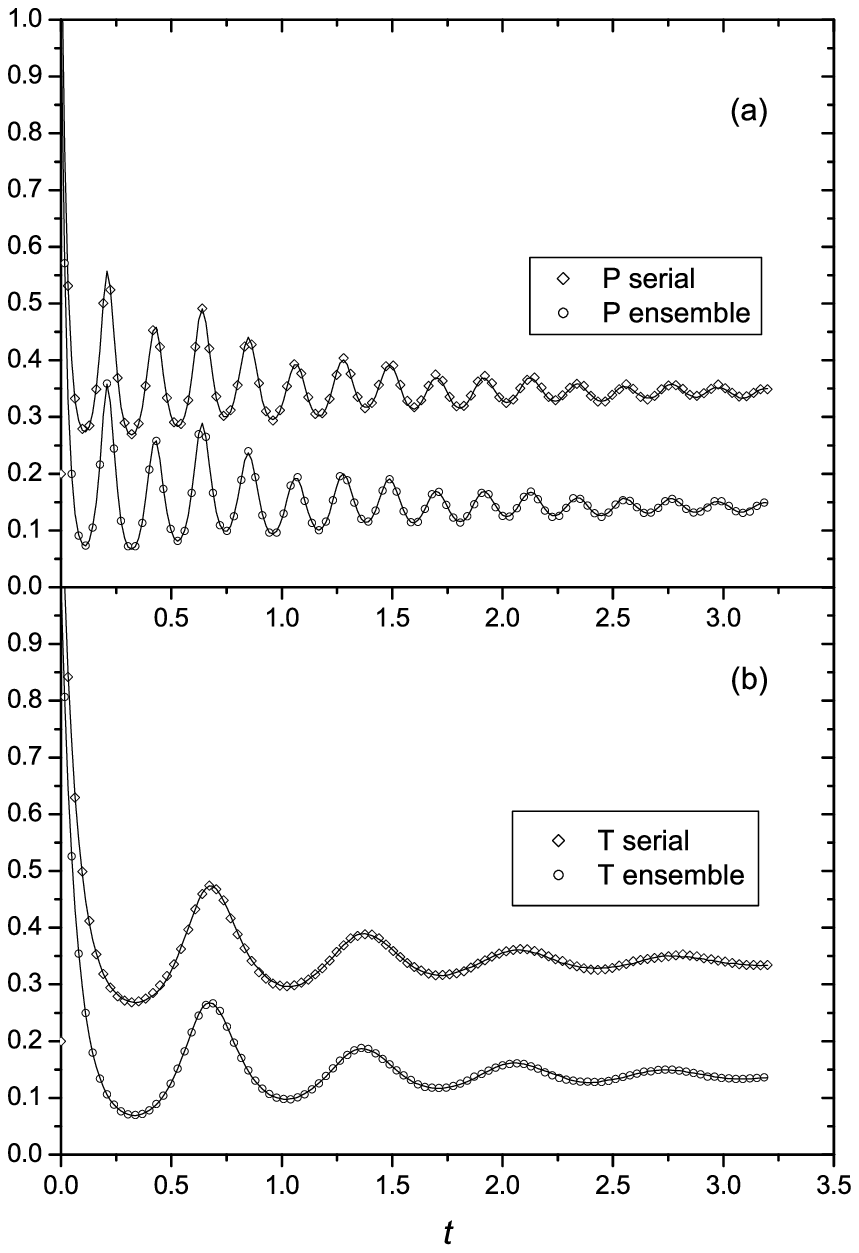}

\caption{Results from applying the fluctuation theorem to the glass for a)
a sudden pressure change, where the symbols are $p(\Delta V>0)/p(\Delta V<0)$
and the solid line is$\left\langle \exp(\beta(P_{2}-P_{1})\Delta V)\right\rangle _{\Delta V<0}$
and b) a sudden temperature change, where the symbols are $p(\Delta I_{E}>0)/p(\Delta I_{E}<0)$
and the solid line is$\left\langle \exp((\beta_{2}-\beta_{1})\Delta I_{E})\right\rangle _{\Delta I_{E}<0}$.
Results from an ensemble of independent initial systems and in addition
from a single initial trajectory (with a time of $t=5$ being computed
between each transient trajectory) are shown for a total of $10^{5}$
pressure or temperature changes. The serial results have been shifted
up, for clarity, by adding 0.2 to the data \label{fig-4}}
\end{figure}
 If we substantially increase the duration the time averages (for
each domain) are constructed on, the peak will be shifted to lower
temperatures as previously shown \cite{Yu-PRE-04}. This requires
the time average to be constructed over some two orders of magnitude
more time than the decay time for the correlation function in Eq.
\ref{correlation1}. This is necessary in order to obtain enough independent
samples for a meaningful estimate of the variance, of the instantaneous
enthalpy appearing in Eq. \ref{Cp-regular}. For a large macroscopic
system we would expect that the specific heat measured over the entire
ensemble would differ very little to that measured from any one of
its members. We are now in a position to make an unambiguous interpretation
of the peak in the specific heat. The peak is observed at the temperature
where the system leaves metastable equilibrium and enters a history
dependent state that requires averages to be computed by Eq. \ref{multi-D-average}
rather than by direct use of Eq. \ref{dist-standard}. The calculation
of both $\left\langle I_{0}\right\rangle _{\alpha}$ and $\left\langle I_{0}^{2}\right\rangle _{\alpha}$
will be different for each domain. If we use time averaging to calculate
these variables on a time scale that falls within the plateau region
for Eq. \ref{correlation1}, see Fig. \ref{fig-2}b, then the amount
of time chosen to form the average is not critical. The peak occurs
because the various ensemble members have become locked in local domains
on the time scale that we are able to access. Near the peak itself
these domains are not expected to be robust.

The multidomain average, Eq. \ref{Cp-concat}, gives the heat capacity
for a glass with robust domains. At the lowest temperatures the heat
capacity reaches the beginning of a plateau, Fig. \ref{fig-3}b. For
the constant volume heat capacity $C_{V}$ this plateau (within uncertainties
due to finite size effects) has a value that is consistent with the
Dulong-Petit law \cite{Landau-Lifshitz-Stat-Mech-1-book}, as would
be expected for an amorphous solid where the potential energy surface
can be modeled as harmonic upon transformation to the orthogonal independent
basis set. This is exactly what we would expect from our local domain
model at low temperatures. 

Testing the integrated transient fluctuation theorem (ITFT) for a
sudden pressure change Eq. \ref{IFT-P} and a sudden temperature change
Eq. \ref{IFT-T} provides further evidence that the Boltzmann distribution
may be used to accurately describe intra-domain statistics in the
glassy state and also that in a properly aged glass, the domains are
robust with respect to the pressure and temperature changes studied
here, Fig. \ref{fig-4}. These equations remain valid whether we subject
an ensemble of simulations (multidomain) to a quench or we sample
from a single trajectory (single domain), which remains stuck in a
single domain. The accuracy with which these relations are satisfied
is powerful independent evidence for the applicability of our assumptions
to the systems studied here. The fact that over the times shown in
Fig. \ref{fig-4}, the ITFT does indeed yield correct results directly
implies that, within experimental tolerance of the data, the phase
space domains must be robust and the number of inter-domain transitions
must be negligible. Unlike the specific heat fluctuation formula this
requires that the domains are robust to finite changes of the state
rather than infinitesimal changes. Thus, given that we have aged the
glass sufficiently that domains are robust the number of transitions
are negligible over the longest time the fluctuation formulae are
computed, we expect Eqs. \ref{IFT-P} \& \ref{IFT-T} to be correct.
If we wished to apply the steady state fluctuation theorem matters
become more difficult \cite{Williams-PRL-06}.

\section{Conclusions}

We have presented a rigorous development of statistical mechanics
and thermodynamics for nonergodic systems where the macroscopic properties
are sensibly time independent and the phase space for the ensemble
is partitioned into robust domains. Using computer simulation we have
carried out various tests on a glassy system and shown that apart
from the immediate vicinity of the glass transition, the computed
results are consistent with our theory. While the intra-domain populations
are individually Boltzmann distributed, the inter-domain populations
are not. 

A correlation function whose decay to zero requires global Boltzmann
weighting, has been derived and it has been shown that it decays on
a reasonable time scale for ergodic systems but not for nonergodic
systems. A second correlation function which decays to zero if the
intra-domain populations are Boltzmann distributed but globally the
inter-domain populations are not, has also been defined. We have developed
expressions for obtaining averages in a multiple domain ensemble and
shown how single domain averages, which always give correct results
in metastable equilibrium, can give spurious results in a history
dependent nonergodic ensemble. The statistical mechanics and thermodynamics
developed here allow the derivation of expressions for multidomain
ensemble averages which give the correct results for time nondissipative
nonequilibrium ensembles. The fundamental origin of the peak in the
specific heat near the glass transition has been unambiguously shown
to be a signature of a transition from metastable equilibrium to a
nonergodic multi-domain ensemble. 

We have shown that the transient fluctuation relations for temperature
and pressure quenches provide independent tests of the fundamental
hypotheses used in our theory: that intra-domain populations are individually
Boltzmann distributed,  that except in the immediate vicinity of the
glass transition the domains are robust with respect to small but
finite variations in thermodynamic state variables, and that the inter-domain
transition rates are negligible. 

%%\bibliographystyle{apsrev}
%%\bibliography{glass}

\begin{thebibliography}{28}
\bibitem[{end()}]{endnote-1}
\bibinfo{note}{An ensemble of 100 simulations was used to obtain the specific
  heat data starting from an initial equilibrium at $T=5$. The ensemble was
  sequentially quenched to the next lower temperature at the completion of each
  cycle. For each cycle the ensemble was equilibrated for a duration of
  $t=2\times 10^3$ before computing a production run of duration $t=1\times
  10^4$. The temperatures were $T=$ 5, 4, 3, 2, 1.75, 1.5, 1.25, 1, 0.9, 0.8,
  0.7, 0.6, 0.55, 0.5, 0.475, 0.45, 0.4375, 0.425, 0.4, 0.35 and 0.3. Some of
  the central difference data in Fig. 3b was computed at different
  temperatures: however all the nonergodic data was computed by strictly
  adhering to the above protocol.}
\end{thebibliography}

\begin{acknowledgments}
We thank the Australian Partnership for Advanced Computing (APAC)
for computational facilities and the Australian Research Council (ARC)
for financial support.
\end{acknowledgments}

\end{document}